*This manuscript is in submission for publication. It has not yet been peer reviewed.*

**Perceptual Ratings Predict Speech Inversion Articulatory Kinematics**

**in Childhood Speech Sound Disorders**

Nina R. Benway[1], Saba Tabatabaee[1], Dongliang Wang[2], Benjamin Munson[3], Jonathan L. Preston[4], Carol Espy-Wilson[1]

*[1]Department of Electrical and Computer Engineering, University of Maryland, College Park, MD*

*[2]Public Health and Preventive Medicine, SUNY Upstate Medical University, Syracuse, NY*

*[3]Department of Speech-Language-Hearing Sciences, University of Minnesota-Twin Cities, Minneapolis, MN*

*[4] Department of Communication Sciences & Disorders, Syracuse University, Syracuse, NY*

**Corresponding author: Nina R Benway** benway@umd.edu, Electrical and Computer Engineering, 380 AV Williams, University of Maryland, College Park, MD 20742

**Funding:** This work was supported by the NIH (T32DC000046-28, N. Benway, Trainee, C. Espy-Wilson, Mentor, and M. Goupell & C. Carr, PIs; R01DC020959, J. Preston, PI) and NSF (BCS2141413, C. Espy-Wilson, PI).

**Abstract**

**Purpose**: This study evaluated whether articulatory kinematics, inferred by Articulatory Phonology speech inversion neural networks, aligned with perceptual ratings of /ɹ/ and /s/ in the speech of children with speech sound disorders.

**Methods**: Articulatory Phonology vocal tract variables were inferred for 5,961 utterances from 118 children and 3 adults, aged 2.25-45 years. Perceptual ratings were standardized using the novel 5-point PERCEPT Rating Scale and training protocol. Two research questions examined if the articulatory patterns of inferred vocal tract variables aligned with the perceptual error category for the phones investigated (e.g., tongue tip is more anterior in dentalized /s/ productions than in correct /s/). A third research question examined if gradient PERCEPT Rating Scale scores predicted articulatory proximity to correct productions.

**Results**: Estimated marginal means from linear mixed models supported 17 of 18 /ɹ/ hypotheses, involving tongue tip and tongue body constrictions. For /s/, estimated marginal means from a second linear mixed model supported 7 of 15 hypotheses, particularly those related to the tongue tip. A third linear mixed model revealed that PERCEPT Rating Scale scores significantly predicted articulatory proximity of errored phones to correct productions.

**Conclusion**: Inferred vocal tract variables differentiated category and magnitude of articulatory errors for /ɹ/, and to a lesser extent for /s/, aligning with perceptual judgments. These findings support the clinical interpretability of speech inversion vocal tract variables and the PERCEPT Rating Scale in quantifying articulatory proximity to the target sound, particularly for /ɹ/.

## Introduction

Childhood speech sound disorder (SSD) reflects neurodevelopmental impairment in speech sound production relative to the normative patterns in a language community (Black et al., 2015). Articulatory kinematic analyses can offer critical insights into the vocal tract configurations underlying speech sound errors in SSD, but traditional articulatory kinematic analyses rely on specialized and often cumbersome instrumentation (Terband et al., 2019). The need for specialized equipment limits the amount of research laboratories doing articulatory kinematic research and reduces diversity of participants represented in these datasets, for a wide range of communication disorders. This limitation raises the question of whether kinematic speech science for children with SSD could be conducted with *acoustic-to-articulatory speech inversion*, the long-studied process of inferring articulatory movements from audio recordings of speech without specialized equipment (e.g., Papcun et al., 1992). While there is evidence that speech inversion inferences provide useful information about speaker-level characteristics, for example, in mental health assessment (Premananth & Espy-Wilson, 2025; Seneviratne & Espy-Wilson, 2022), there is not yet evidence that speech inversion provides clinically interpretable articulatory information about the segmental aspects of errored speech in SSD. This study investigates that possibility.

### Speech sound disorders and phonetic speech sound errors

The speech errors of SSD often pattern differently based on a child's age. Phonetic transcriptions of the speech of preschool-aged children with SSD suggests that their errors are categorical speech sound deletions (i.e., producing boat as [boʊ]) or speech sound substitutions (i.e., producing see as [ti]). However, when these errors continue into adolescence (Lewis et al., 2015) or adulthood (Flipsen, 2015), the errors are often transcribed as phonetic deviations from typical articulatory patterns rather than as phonemic deletions or substitutions. Phonetic deviations in American English most frequently impact rhotics (i.e., /ɹ/) and sibilants (e.g., /s/). These speech sound errors include productions of /ɹ/ or /s/ that are perceptually distinct from /ɹ/ or /s/, but do not fully resemble other sounds in the language.

*American English /ɹ/*

Production of fully rhotic American English /ɹ/ involves a complex vocal tract configuration. Either *bunched* or *retroflexed* vocal tract configurations yield fully rhotic /ɹ/, and these two canonical configurations can be considered endpoints of an /ɹ/ tongue shape continuum (Boyce, 2015). Both configurations share features like pharyngeal narrowing and lateral bracing of the tongue against the upper molars, but differ in tongue tip elevation relative to the tongue blade. Either bunched and retroflexed configurations can result in the salient acoustic features of a fully rhotic /ɹ/, notably a low third formant that is close in frequency to the second formant (Delattre & Freeman, 1968; Espy-Wilson et al., 2000). The characteristic low third formant of fully rhotic American English is associated with the narrowing of the palatal constriction (Harper et al., 2020). "Derhotic" /ɹ/ productions differ enough in vocal tract configuration from adult rhotics to be perceived by listeners as non-adult-like. For derhotic /ɹ/, midsagittal ultrasound imaging of the tongue during speech may reveal (1) a too-low tongue tip or blade, (2) a tongue dorsum that is too high or too far back, and/or (3) limited tongue root retraction (Preston, Benway, et al., 2020). Coronal ultrasound imaging may also reveal insufficient lateral bracing of the tongue against the upper molars. Perceptually, derhotic /ɹ/ may often have minimal, moderate, or significant characteristics of other sounds (e.g., Ball, 2017). These characteristics may reflect voiced approximants (i.e., labiovelar [w], labiodental approximant [ʋ], alveolar lateral [l], or velar [ɰ]), fricatives (i.e., velar [ɣ] or uvular [ʁ]), and/or vowels (i.e., open-mid back unrounded [ʌ], near-close near-back rounded [ʊ], or open-mid back rounded [ɔ]).

This study uses speech inversion to kinematically describe the articulation of derhotic /ɹ/ productions in individuals with SSD, in comparison to fully rhotic /ɹ/. Specifically for /ɹ/, the study examines two clinically relevant perceptual error subtypes: derhotic /ɹ/ → [w] phones and derhotic /ɹ/ → [+vocalic] phones. Productions of phonemic /w/ and /ʌ/ were included for comparison to /ɹ/ as positive controls. Phonemic /w/ was paired to derhotic /ɹ/ → [w] phones, and phonemic /ʌ/ was paired to derhotic /ɹ/ → [+vocalic][1] phones.

In this context, a positive control refers to a phoneme with a well-established articulatory configuration providing a reliable reference point for evaluating the accuracy and interpretability of speech inversion results. The rationale for testing speech inversion patterns among positive control phonemes was that their articulatory configurations are well established, whereas the error subtypes identified in this study are based on perceptual judgments due to the lack of a suitable ground-truth articulography dataset for SSD. From the viewpoint that, for example, the phoneme /w/ could be seen as the "full substitution" endpoint of derhotic /ɹ/ → [w] productions, positive control phonemes could help interpret unexpected inversion patterns and distinguish whether such results arise from inversion limitations or from perceptual labeling.

*American English /s/*

Despite the immense variation in American English /s/ production, /s/ is generally articulated at the alveolar ridge, where a constriction between the tongue and the alveolar ridge creates a narrow passage for airflow. This narrow passage is reinforced by lateral bracing of the tongue against the upper molars. The part of the tongue closest to the alveolar ridge may represent the tongue tip, in tip-up apico-alveolar configurations, or tongue blade, in tip-down lamino-alveolar configurations (Dart, 1998; Narayanan et al., 1995). In either configuration, the characteristic stridency of /s/ arises from the high-velocity turbulent airflow that is channeled through this constriction and subsequently directed against the incisors (Stevens, 1998). The interaction between the airflow and the anterior cavity formed by the tongue, upper teeth, and alveolar ridge enhances high-frequency energy, contributing to the perceptually distinct, hissing quality of /s/. For /s/ speech errors, midsagittal ultrasound imaging of the tongue during speech may reveal anterior/posterior tongue placement, while coronal ultrasound imaging may reveal insufficient lateral bracing of the tongue against the upper molars. Subtle differences in /s/ articulation can impact the acoustic spectrum of /s/ (Munson, 2004). Perceptual characteristics of /s/ misarticulation may resemble the perceptual characteristics of other coronal fricatives such as the voiceless dental /θ/, voiceless alveolar lateral /ɬ/, or voiceless postalveolar /ʃ/ (e.g., Daniloff et al., 1980).

---

[1] This paper uses "/ɹ/ → [+vocalic]" instead of "/ɹ/ → [ʌ]" because the perceptual label for the rating scale described in the Methods section did not specify among possible vowel substitutions.

In addition to /ɹ/, this study also kinematically describes the articulation of clinically incorrect /s/ in comparison to clinically correct /s/ using speech inversion. For /s/, the study examines three clinically relevant /s/ error subtypes: fronted/dentalized /s/ → [s̪] errors, backed/palatalized /s/ → [sʲ] errors, and lateralized /s/ → [sˡ] errors. Productions of phonemic /θ/, /ʃ/, and /l/ were also included as positive controls. Phonemic /θ/ was paired with dentalized /s/ → [s̪] phones, and phonemic /ʃ/ was paired with palatalized /s/ → [sʲ] phones. Because the voiceless alveolar lateral fricative /ɬ/ is neither phonemic nor allophonic in American English, this study uses the voiced alveolar lateral approximant /l/ as a positive control for lateralized /s/ → [sˡ] phones.

**Automating articulatory kinematic analyses with acoustic-to-articulatory speech inversion**

Speech inversion is the long-studied process of inferring articulatory movements from a continuous speech signal, allowing researchers to approximate kinematic analysis of speech in datasets that were not collected through traditional kinematic instrumentation (e.g., Papcun et al., 1992). Modern speech inversion approaches this task through neural networks that are trained to associate ground-truth kinematic data with coregistered speech audio. The kinematic data used to train speech inversion neural networks is often obtained through x-ray microbeam or electromagnetic articulography systems that glue sensors to the lips and tongue in order to track articulation, as described by Westbury et al. (1994) and Rebernik et al. (2021).

Our team's Speech Inversion systems (e.g., Siriwardena et al., 2024; Siriwardena & Espy-Wilson, 2023), predict the *constriction location* and *constriction degree* of *vocal tract variables* from speech using the Articulatory Phonology framework (Browman & Goldstein, 1992). Articulatory Phonology-based speech inversion transforms the absolute (X,Y) location of ground-truth articulography sensors, in millimeters, into normalized *vocal tract variables* for neural network training and inference. For example, instead of using the tongue tip sensor's exact (X,Y) position as ground truth, the systems calculates *tongue tip constriction location* and *tongue tip constriction degree* from the articulography data. These normalized *vocal tract variable* representations generalize better across speakers than absolute articulography (X-Y) sensor coordinates, which vary with anatomical differences in vocal tract size (Mitra et al., 2010). Articulatory Phonology-based systems offer particular advantages for child speech research because the normalization facilitates the application of adult-trained models to child speech, mitigating the lack of publicly available child speech inversion systems (cf: Oohashi et al., 2017; Serkhane et al., 2007, discussed below). This normalization also supports speaker-independent speech inversion, which has particular utility for future clinical speech applications. Speaker-dependent systems perform well, but their reported accuracy in predicting the kinematic ground truth will likely generalize most to future speakers with sensor data, which is a barrier for future clinical use.

Articulatory Phonology models the vocal tract with six *vocal tract variables*: *lip aperture*, *lip protrusion*, *tongue tip constriction location*, *tongue tip constriction degree*, *tongue body constriction location*, and *tongue body constriction degree*. Recent speech inversion research has demonstrated that also incorporating glottal *source variables* (*fundamental frequency, periodic energy,* and *aperiodic energy*) significantly enhances the accuracy of vocal tract gesture estimation (Deshmukh et al., 2005;

Siriwardena & Espy-Wilson, 2023). *Source variables* may improve speech inversion by providing timing information that constrains the estimation of dynamic articulatory gestures, while also offering complementary cues regarding nonlinear interactions between the glottal source and the vocal tract filter.

*Geometric transformations for Articulatory Phonology vocal tract variables*

Attia et al. (2024) describe the calculation of Articulatory Phonology *vocal tract variables* from (X,Y) sensor coordinates in the Wisconsin Xray microbeam dataset (Westbury et al., 1994), the dataset serving as the kinematic ground truth for our team's speech inversion systems. Figure 1 shows the location of the untransformed sensor (i.e., gold pellet) locations on the lips and tongue as well as the transformed *vocal tract variables*. The T1 pellet represents the ventral tongue, ($\bar{x} = 8.5$ mm from the tongue apex, $x_{\sigma} = 1.07$ mm), T2 represents the mid-ventral tongue ($\bar{x} = 25.2$ mm from the tongue apex, $x_{\sigma} = 2.44$ mm), T3 represents the mid-dorsal tongue ($\bar{x} = 43.8$ mm from the tongue apex, $x_{\sigma} = 3.49$ mm), and T4 represents the dorsal tongue ($\bar{x} = 60.1$ mm from tongue apex, $x_{\sigma} = 4.14$ mm). The UL pellet represents the upper lip, and the LL pellet represents the lower lip. There is also an incisor reference pellet. For the tongue body (Figure 1 panel A), *tongue body constriction degree* is calculated as the minimum distance between an arc connecting pellets T2, T3, T4, and the trace of the maxilla/velum/anterior pharyngeal wall. *Tongue body constriction location* is calculated as the horizontal offset of the T2-T4 arc from the incisor origin at the point of narrowest *tongue body constriction degree*. For the tongue tip (Figure 1 panel B), *tongue tip constriction degree* is calculated as the minimum distance between pellet T1 and the maxilla trace. *Tongue tip constriction location* is calculated as the horizontal offset of pellet T1 from the incisor origin at the point of narrowest *tongue tip constriction degree*. For the lips (Figure 1 panel C), LA is calculated as the Euclidean distance between pellets UL and LL, and LP is calculated as the horizontal offset of LL and the incisor origin. To aid the interpretation of *vocal tract variables* in the present study against an articulatory phonetic space, we inverted the signs of tongue tract variables relative to Attia et al. (2024), so positive tongue variable values would reflect more anterior articulation and higher (i.e., more constricted) articulations. The present study examines inferred *vocal tract variables* operationalized according to these geometric transformations. Details regarding the training of the speech inversion neural network are presented in Methods.

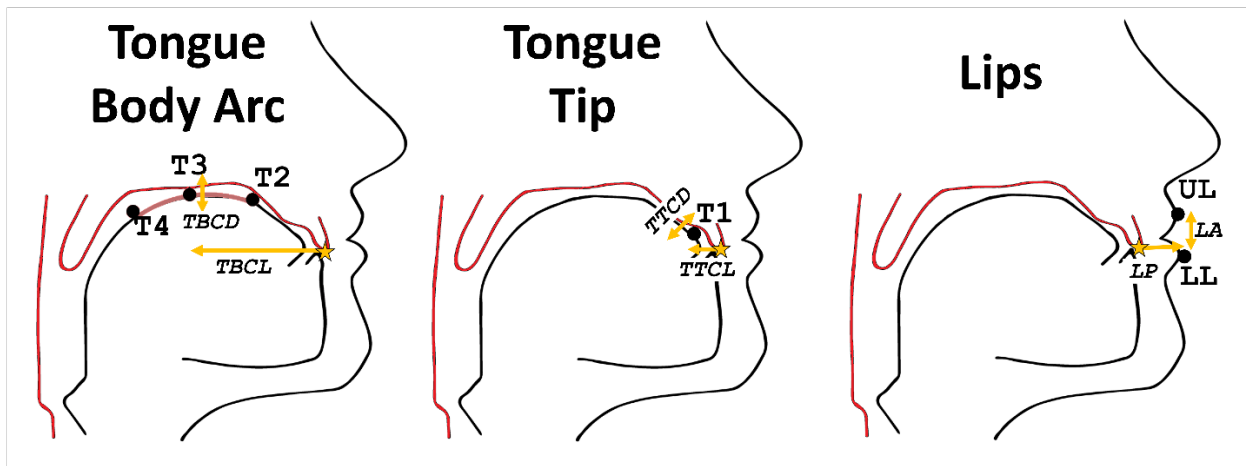

Figure 1: *Converting articulography data to Tract Variables.* The geometric transformations of upper lip (UL), lower lip (LL), tongue tip (T1), and tongue body (T2-T4) Wisconsin X-ray microbeam pellets to Articulatory Phonology Tract Variables lip aperture (LA), lip protrusion (LP), tongue tip constriction location (TTCL), tongue tip constriction degree (TTCD), tongue body constriction location (TBCL), and tongue body constriction degree (TBCD). The star on the tip of the incisors indicates the origin from which constriction location offsets are calculated, and the point that minimizes tongue tip or tongue body distance from the palatal trace.

*Use of adult-trained speech inversion for child speech*

The current lack of suitable public data to train a de novo child speech inversion system has led child speech scientists to rely on adult-trained speech inversion systems. Serkhane et al. (2007) and Oohashi et al. (2017) used an adult articulatory model based on 519 midsagittal cineradiographic frames from one adult French speaker (Maeda, 1990). Because the articulatory model was parametric, the articulatory parameters could be scaled by anatomical growth factors derived from cineradiographic data on children's vocal tract development (Goldstein, 1980; Ménard et al., 2004). Serkhane et al. (2007) inverted this articulatory model to compare articulatory configurations in infant vocalizations at four and seven months of age, showing articulatory exploration and the role of the jaw increased with development. Oohashi et al. (2017) inverted this same model to examine vowel-like sounds produced by three Japanese children aged 6 to 60 months, providing evidence that vowel production involves gradual functionalization of articulatory parameters. In both studies, the formant frequencies for vocalizations and vowels of interest were transformed into articulatory parameters to estimate developmental changes in of jaw articulation, tongue dorsum position, tongue dorsum shape, tongue apex position, lip aperture, lip protrusion, and larynx height.

Benway et al. (2023) used inferences from an adult-trained speech inversion neural network (Siriwardena & Espy-Wilson, 2023) to improve performance of a bidirectional long short-term memory machine learning classifier when predicting clinician judgments of rhoticity, outperforming a previous formant-based system (Benway, Preston, Salekin, et al., 2024). Inferred *tongue body constriction location* significantly differed between utterances perceived as fully rhotic and derhotic. In a separate study, Benway, Preston and Espy-Wilson (2024) showed that vocal tract gesture coordination values derived from adult-trained inversion inference showed high performance when classifying the speech of 96 children as young as 7 years with two subtypes of SSD: childhood apraxia of speech and non-apraxia SSD. Ablating *source variables* from the classifier significantly impaired classifier performance, implicating source–filter coordination in distinguishing these SSD subtypes. The importance of source variables replicated prior perceptual findings in the dataset  (Benway & Preston, 2020), further demonstrating the sensitivity of adult-trained inversion to child speech in SSD. In the continued absence of publicly available child articulography datasets, these studies show that adult-trained speech inversion models capture meaningful articulatory information from child speech.

**Purpose, research questions, and *a priori* hypotheses**

*Overview of research questions*

This study is a speech science investigation of articulatory patterns among speech sound productions from children with SSDs. The purpose of the study was to examine how Articulatory Phonology *vocal tract variables*, inferred with acoustic-to-articulatory speech inversion neural networks, matched the clinical perception of /ɹ/ and /s/ errors subtypes in children with SSD. We examined three research questions concerning /ɹ/ and /s/, the most commonly errored speech sounds in chronic SSDs.

Our first, preliminary, research question asked if adult-trained speech inversion *vocal tract variables* were sensitive to phonemic articulatory differences in child speech for the positive control phonemes versus correct /ɹ/ or /s/. For example, we examined how speech inversion *vocal tract variables* for /w/ differ from /ɹ/. Our secondary research question asked whether adult-trained speech inversion *vocal tract variables* were sensitive to phonetic articulatory differences in child SSD, showing similar directionality in tract variable differences as a paired positive control phoneme. For example, our second research question examined how speech inversion *vocal tract variables* for derhotic /ɹ/ → [w] phones differ from both correct /ɹ/ and positive control /w/. We constructed 33 *a priori* articulatory hypotheses for these first and second research questions, which we step through in the following section. An illustrative hypothesis, relating to the examples above, is that we expected that the positive control phoneme /w/ would show greater lip protrusion and a more posterior tongue body constriction location compared to phonemic /ɹ/, and that derhotic /ɹ/ → [w] phones would show similar but intermediate labiovelar patterns, consistent with viewing derhotic /ɹ/ → [w] phones along a continuum of /r/ to /w/. Each of the 33 hypotheses were set before the analysis was run and were constructed based on our team's experience with articulatory phonetics and acoustic phonetics across languages (International Phonetic Association, 1999; Li & Munson, 2016; Munson, 2004; Munson & Coyne, 2010), ultrasound imaging of the tongue in SSD (Benway et al., 2021; Preston, Hitchcock, et al., 2020; Preston et al., 2017; Preston, McAllister, et al., 2019; Preston, McCabe, et al., 2019), magnetic resonance imaging of the vocal tract (Zhou et al., 2008), vocal tract modeling (Espy-Wilson et al., 2000; Zhou et al., 2007), and Articulatory Phonology *vocal tract variables* (Attia et al., 2024; Mitra et al., 2010).

While our first and second research questions investigated *categorical* differences between phones of interest, our third research question investigated if the *degree* of articulatory and perceptual errors were systematically related. Specifically, our third research question explored if scores from the PERCEPT Rating Scale, a researcher-developed 5-point scale and training pipeline capturing perceptual variation in phoneme accuracy, predicted the articulatory proximity of the errored phone to the correct target. We hypothesized that as PERCEPT Rating Scale score decreased (i.e., were rated as more incorrect), articulatory distance from the correct *vocal tract variable* configuration would increase. This analysis complements the first two research questions by testing whether *vocal tract variables* can capture the perceptual magnitude of speech sound errors in /ɹ/ and /s/, which is highly relevant for clinical assessment and treatment monitoring, while also providing additional evidence for the clinical and articulatory validity of speech inversion *vocal tract variables* for children with SSD.

*Individual articulatory hypotheses*

We constructed 33 hypotheses that related to children's productions of /ɹ/ and /s/ for our first and second research questions: 11 for the positive control phonemes compared to the correct target, 11 for the phonetic error compared to the correct target, and 11 for the positive control phoneme compared to the paired phonetic error. These hypotheses are summarized in Table 1 in both articulatory and *vocal tract variable* terms. Notably, we only considered hypotheses that were reasonably robust to coarticulation. For example, we could reasonably expect that *lip protrusion* is greater in /ʃ/ and palatalized [sʲ] than in /s/, all else being equal. However, that hypothesis was not tested, because it was beyond the scope of the present study to examine the coarticulatory distribution of biphones in the dataset.

Table 1: *Hypotheses for the first and second research questions.*

| Target Phoneme | Positive Control Phoneme | Perceived Phonetic Error | Articulatory Hypothesis | *Vocal Tract Variable* Interpretation of Hypothesis |
|---|---|---|---|---|
| /ɹ/ | /w/ | /ɹ/ → [w] | Lip are more protruded than correct /ɹ/ | LP is positive compared to correct /ɹ/ |
| | | | Tongue tip/blade is lower than correct /ɹ/ | TTCD is negative compared to correct /ɹ/ |
| | | | Most constricted part of tongue body is more posterior than correct /ɹ/ | TBCL is negative compared to correct /ɹ/ |
| /ɹ/ | /ʌ/ | /ɹ/ → [+vocalic] | Lip are less protruded than correct /ɹ/ | LP is negative compared to correct /ɹ/ |
| | | | Tongue tip/blade is lower than correct /ɹ/ | TTCD is negative compared to correct /ɹ/ |
| | | | Tongue body constriction is looser than correct /ɹ/ | TBCD is negative compared to correct /ɹ/ |
| /s/ | /θ/ | /s/ → [s̪] | Tongue tip/blade is more anterior than correct /s/ | TTCL is positive compared to correct /s/ |
| | | | Tongue body constriction is looser than correct /s/ | TBCD is negative compared to correct /s/ |
| /s/ | /ʃ/ | /s/ → [sʲ] | Tongue tip/blade is more posterior than correct /s/ | TTCL is negative compared to correct /s/ |
| /s/ | /l/ | /s/ → [sˡ] | Tongue tip/blade is higher than correct /s/ | TTCD is positive compared to correct /s/ |
| | | | Tongue body constriction is looser than correct /s/ | TBCD is negative compared to correct /s/ |

*Note*: LP = lip protrusion, LA = lip aperture, TTCL = tongue tip constriction location, TTCD = tongue tip constriction degree, TBCL = tongue body constriction location, TBCD = tongue body constriction degree, [s̪] = dentalized /s/, [sʲ] = palatalized /s/, and [sˡ] = lateralized /s/.

For /ɹ/, the positive control phoneme /w/ could be seen as the "full substitution" endpoint of derhotic /ɹ/ → [w] errors. For the first research question, we hypothesized that phonemic /w/ would have more lip protrusion, a lower tongue tip/blade, and a more posterior tongue body than correct /ɹ/, reflecting that /w/ is a labiovelar sound. For the second research question, we hypothesized that derhotic /ɹ/ → [w] phones would show similar articulatory patterns compared to /w/, but of an intermediate degree.

Similarly, the positive control phoneme /ʌ/ can be seen as a "full substitution" endpoint of derhotic /ɹ/ → [+vocalic] errors. For the first research question, we hypothesized that /ʌ/ would have less lip protrusion, a lower tongue tip/blade, and a looser tongue constriction than correct /ɹ/, reflecting the neutral vocal tract configuration of /ʌ/. For the second research question, we hypothesized that derhotic /ɹ/ → [+vocalic] phones would show similar articulatory patterns compared to /ʌ/, but of an intermediate degree.

For /s/, the positive control phoneme /θ/ can be seen as the "full substitution" endpoint of dentalized /s/ → [s̪] errors. For the first research question, we hypothesized that /θ/ would have a tongue tip/blade that was more dental/anterior and a tongue body constriction that was looser than correct /s/, reflecting the dental place of articulation for /θ/ and the lateral lingual bracing of /s/ (even though the tongue body is not considered a primary articulator for /s/). For the second research question, we hypothesized that dentalized /s/ → [s̪] phones would show similar articulatory patterns compared to /θ/, but of an intermediate degree.

Similarly, the positive control phoneme /ʃ/ can be seen as the "full substitution" endpoint of palatalized /s/ → [sʲ] errors. For the first research question, we hypothesized that /ʃ/ would have a tongue tip/blade that was more posterior than correct /s/, reflecting the postalveolar place of articulation for /ʃ/. For the second research question, we hypothesized that errored /s/ → [sʲ] phones would show similar articulatory patterns compared to /ʃ/, but of an intermediate degree.

Finally, we also wanted to examine lateralized /s/ → [sˡ] errors because of their clinical relevance, as these errors are common in older children with SSD. However, speech inversion provides information based on midsagittal speech movements, and we were curious about the extent to which speech inversion would capture lateral tongue dynamics for this fricative (and other sounds for which lateral movement is relevant). Moreover, a "full substitution" endpoint for lateralized /s/ would ideally be /ɬ/, which is neither phonemic nor allophonic in American English. For this reason, we selected (word-initial, alveolar) /l/ as the best fit for a positive control and endpoint for lateralized /s/ errors. For the first research question, we hypothesized that /l/ would have a higher tongue tip/blade and a looser tongue body constriction than correct alveolar /s/, reflecting more complete alveolar contact in /s/ and lack of lateral lingual bracing for lateral /l/. For the second research question, we hypothesized that errored /s/ → [sˡ] phones would show similar, intermediate articulatory patterns as /l/.

## Methods

### Dataset

This project analyzed speech audio data from 17,083 candidate audio files recorded from 118 children ($\bar{x} = 9.56, \sigma = 2.96$) and 3 adults ($\bar{x} = 28.3, \sigma = 14.5$) aged 2 years 3 months to 45 years. several IRB-approved clinical trials, longitudinal studies, and cross-sectional studies (Benway & Preston, 2024; Herbst et al., 2025; Preston et al., 2024; Preston & Edwards, 2007; Preston, Hitchcock, et al., 2020; Preston et al., 2013; Preston et al., 2014; Sjolie et al., 2016; Wong et al., 2025). Utterances in each of these projects were recorded through wordlist recordings or during speech therapy activities. All audio was recorded as 44.1 kHz lossless PCM 16-bit audio in .wav containers. Linguistic stimuli are comprised of words and phrases, such as "rabbit" and "ruling the kingdom" (for /ɹ/), and "jacks" and "salad" (for /s/). Each utterance contained only one target /ɹ/ or /s/.

*Independent variables for research questions: PERCEPT Rating Scale Error Labels and Scores.*

The speech data have been labeled with expert clinician perceptual ratings of the target /ɹ/ or /s/ using custom rating scales running as a private web app, the PERCEPT Rating Scale (Figure 2). Similar perceptual rating scales, in 5-point Likert format, were used for /ɹ/ and /s/. Three points on the scales indicated clinically incorrect productions: scale points 1 (substitution/omission), 2 (major distortion), and 3 (close, leaning incorrect/minor distortion). Two points on the scales indicated clinically correct productions: scale points 4 (close, leaning correct) and 5 (fully correct). All productions marked as scale point 1 (substitution/omission) required a secondary selection indicating the categorical label of the speech error subtype. This label was optional for scale points 2 and 3 (distortions). For /ɹ/, possible speech error subtypes included: */w/-error*, */l/-error*, *vowel error*, *omitted*, and *other*. For /s/, possible speech error subtypes included: *dentalized*, *lateralized*, *palatalized*, *affricate*, *stopped*, *omitted*, and *other*. Raters were masked to participant information and timepoint of audio collection (e.g., pre-treatment, post-treatment). Audio from different speakers were randomly mixed into batches of 100 files. Raters also had the opportunity to mark files that contained unusable audio, incomplete audio, or transcript mismatches between the orthographic label for the file and the audio content of the file.

Figure 2: *PERCEPT Rating Scale*. The PERCEPT Rating Scale web app, a Likert scale for the perceptual rating of sounds in the context of speech sound disorder. This example shows the scale for /s/ during a training module. Also visible are web app utilities for playing the file and entering comments. The three leftmost scale points correspond to clinically incorrect ratings and the two rightmost scale points correspond to clinically correct ratings. Possible error subtype labels are shown in the bottom left.

The audio files were rated by eight licensed speech-language clinicians associated with the Syracuse University Speech Production Lab. Raters were calibrated to the PERCEPT Rating Scale through a dedicated pipeline consisting of pre-training activities and four training modules per target sound. The audio examples selected for the training pipeline had received a categorical label and score reflecting the agreement of the study's first, fourth, and fifth authors (i.e., the "expert panel"). Pre-training activities provided detailed written descriptions of the scale points and error subtypes along with audio examples. Each of the four training modules resembled the rating task; raters heard an audio example and indicated their rating using the PERCEPT Rating Scale webapp. Raters advanced to the next training module in the pipeline when both a binary criterion and a continuous criterion were met. The binary criterion required a rater to have ≥ 90% agreement with the expert panel for the "correct" and "incorrect" categories that were formed by collapsing scale scores 1-3 and 4-5. The continuous criterion required that ≥ 90% of the 5-point ratings must be within one scale point of the expert panel consensus score. Raters who did not achieve both criteria during a training module could self-study the audio examples and re-take the module until passing the criteria for that module. Raters completed the training after having passed three 25-item modules and one 50-item module. The rater training pipeline was done separately for each target sound /ɹ/ and /s/, after which raters moved on to rate audio for this study and other studies associated with R01DC020959 (J. Preston, PI). The training specified that raters pass a 50-item maintenance module after every 20 modules, so raters could recalibrate against internal perceptual drift; however, all audio for this study was rated within the first 20 modules after training for all raters.

A total of 56,929 perceptual ratings were analyzed for this study, from the 17,083 candidate audio files. Several raters rated each file because non-perfect agreement is known to occur even among expert clinician raters (Klein et al., 2013). 1,972 of the 17,083 audio files were excluded from the analysis because raters indicated issues that would impact forced alignment success (e.g., unusable audio, partial audio, transcript mismatch, or target sound omission). This study examines the remaining utterances that met the criteria for either *fully correct target phones* or *labeled non-correct phones*: Correct utterances were defined by unanimous rater agreement that the /ɹ/ or /s/ in the audio was fully correct. Non-correct phones were defined as utterances receiving average score less than 5 that had a consensus among all raters on the categorical error subtype label for an audio file. Consensus was defined as at least two raters agreeing on the specific error subtype label, with the agreeing raters also representing the majority of the total raters for that file. Ground truth error subtypes with fewer than 15 utterances were excluded from the analysis (/ɹ/ → [l] errors, "other" /ɹ/ errors, "other" /s/ errors, /s/ → [t] errors, and /s/ → [ʧ] errors). These criteria retained 5,961 audio files in the analysis: 3,256 for /ɹ/ and 2,431 for /s/.

The 17,083 candidate audio files analyzed for this study were also searched for positive control phonemes /w/, /ʌ/, /θ/, /ʃ/, or /l/. These phonemes were assumed to be correctly articulated. This search added 429 /w/ phonemes (cf., /ɹ/ → [w] phones), 1,002 /ʌ/ phonemes (cf., /ɹ/ → [+vocalic] phones), 687 /θ/ phonemes (cf., /s/ → [s̪] phones), 636 /ʃ/ phonemes (cf., /s/ → [sʲ] phones), 167 word-initial /l/ phonemes (cf., /s/ → [sˡ] phones) to the analysis. Details about correct /ɹ/ and /ɹ/ error subtypes, correct /s/ and /s/ error subtypes, and positive control phonemes are shown in Table 2. These *phones* were included as an independent variable in the statistical analysis.

Table 2: *Phones of interest analyzed in the study.*

| Phoneme Target | Phone | PRS Score (Mean, STD) | Count of Phones |
|---|---|---|---|
|  | Correct /ɹ/ | 5 (0) | 1,719 |
|  | /ɹ/ → [w] | 1.5 (.6) | 591 |
| /ɹ/ | /w/ | N/A | 429 |
|  | /ɹ/ → [+vocalic] | 1.5 (.6) | 946 |
|  | /ʌ/ | N/A | 1,002 |
|  | Correct /s/ | 5 (0) | 1,401 |
|  | /s/ → [s̪] | 2.4 (.8) | 604 |
|  | /θ/ | N/A | 687 |
| /s/ | /s/ → [sʲ] | 2.6 (.9) | 154 |
|  | /ʃ/ | N/A | 636 |
|  | /s/ → [sˡ] | 2.2 (.7) | 272 |
|  | /l/ | N/A | 167 |

*Note*: N/A entries reflect that positive control phonemes were assumed to be correctly produced. PRS = PERCEPT Rating Scale, STD = standard deviation

*Phone segmentation boundary alignment*

For this study, we only wanted to evaluate the portion of the speech inversion signal associated with the /ɹ/, /s/, or positive control phonemes within the audio files. To do this, we placed phone segmentation boundaries with the Montreal Forced Aligner version 3.0.2, using our custom PERCEPT-TX child speech acoustic models for alignment. The rationale for using custom acoustic models is that target sound in speech therapy audio recordings is often elongated compared to conversation speech (i.e., "rrrrrrrrrrrabbit"), and our preliminary work has shown that adapting the default Librispeech adult acoustic models to these speech patterns improves forced alignment for the target sound compared to non-adapted adult Librispeech acoustic models. The start and end samples identified through forced alignment for each target phone were used to extract the corresponding segment of the speech inversion *vocal tract variables*.

**Speech Inversion Neural Network**

The speech inversion neural network was trained for this study, extending a baseline HuBERT-Large model reported by Attia et al. (2024). For the purposes of *training* the speech inversion neural network, *vocal tract variable* values were transformed from the adult articulography recordings in the Wisconsin X-ray Microbeam dataset to serve as the neural network's ground truth. The neural network was trained to infer the values of *vocal tract variable* values from the adult audio. Then, the trained speech inversion system was used to infer the *vocal tract variable* values for the child SSD audio. Because no child articulography ground truth was available for fine-tuning, the adult-trained model was applied directly to the child SSD audio. The inferred *vocal tract variable values* for the child SSD speech served as the dependent variables in the statistical analyses testing the research questions.

*Training data*

Training data consisted of the speech and articulatory movements of 46 adult speakers (21 male

and 25 female) from the University of Wisconsin X-ray Microbeam dataset (Westbury et al., 1994). The

Wisconsin X-ray Microbeam dataset contains synchronized audio recordings and time-series data of the

(X-Y) positions of gold pellets attached to the speaker during speech. A total of 5.3 hours of recordings

were available for training after reconstructing 1.3 hours of mistraced data in the same manner as Attia

et al. (2024). The dataset splits for this study used the same 36 speakers in the training set, the same 5

speakers in the validation set, and the same 5 speakers in the test set as Attia et al. (2024) to facilitate a direct comparison of results. No one speaker showed up in multiple dataset splits to promote a speaker independent speech inversion system.

*Ground truth: Articulatory Phonology vocal tract variables*

Recall that the Articulatory Phonology *vocal tract variables* are *lip aperture*, *lip protrusion*,

*tongue tip constriction location*, *tongue tip constriction degree*, *tongue body constriction location*, and

*tongue body constriction degree*. The original (X-Y) pellet coordinates in the Wisconsin X-ray

Microbeam dataset were converted into the six Articulatory Phonology *vocal tract variables* using the

geometric transformations described prior. The *vocal tract variables* were subsequently normalized to

the range of [-1, 1] using speaker-level min-max normalization as in Equation 1. We also calculated

glottal *source variables,* fundamental frequency, periodic energy, and aperiodic energy, using the APP

detector algorithm (Deshmukh et al., 2005).

$$\text{Normalized TV} = 2 * \frac{\text{TV} - \text{Min (TV)}}{\text{Max(TV)} - \text{Min (TV)}} - 1$$

(Equation 1)

*Feature input: WavLM audio embeddings*

Our speech inversion system leveraged the WavLM-Large model (Chen et al., 2022) as a pretrained feature extractor to generate self-supervised embeddings from the Wisconsin X-ray Microbeam speech audio. These embeddings served as input to the neural network. Like HuBERT-Large (Hsu et al., 2021), WavLM-Large is a transformer-based model trained on large-scale audio corpora using masked speech predictions. Such models produce hierarchical representations of speech by learning to reconstruct windows of the speech signal that were masked from the model. The use of contrastive loss during reconstruction encourages the model to reconstruct masked windows of the speech signal that are maximally different from neighboring windows, which teaches the model to reproduce speech context that is maximally salient within the masked window. This results in representations that capture the most salient acoustic, phonetic, and suprasegmental patterns from the audio at a given time step. WavLM-Large extends the HuBERT-Large architecture by incorporating a multi-task training objective that includes speech denoising and speaker-aware pretraining, which was found to improve performance on automated speech analysis tasks (Chen et al., 2022).

Specifically, the WavLM-Large hierarchical embeddings derived from the Wisconsin X-ray Microbeam dataset audio were organized as 25-layer embeddings extracted at the default 50 Hz frame rate. These embeddings provided the input features for the speech inversion network's convolutional and recurrent layers.

*Neural network architecture*

The speech inversion neural network architecture is shown in Figure 3. The self-supervised embeddings of the Wisconsin X-ray Microbeam dataset audio were processed through a series of convolutional and recurrent layers to capture local temporal and spectral relationships in the input. The first convolutional layer reduced the 25-layer representation to 16 layers, followed by a second convolutional layer that reduced the representation to a single layer. This single-layer representation was then passed through two gated recurrent unit layers that were designed to capture temporal dependencies in the speech signal. The first gated recurrent unit layer had 256 units, followed by a second gated recurrent unit layer with 128 units. The output of the gated recurrent unit layer was upsampled by a factor of two to match the 100 Hz target sampling rate for speech inversion. Finally, the upsampled features were passed through two dense layers: the first with 128 units and the second with 9 units. The last dense layer served as the mapping function from latent acoustic representations to the nine parameters inferred by the model (i.e., six vocal tract variables, and three glottal source variables).

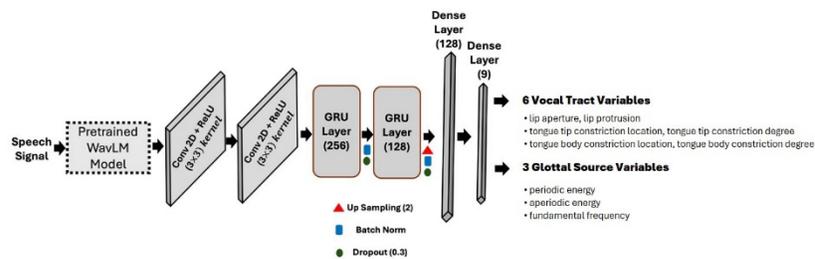

Figure 3: *Neural network architecture for speech inversion.* The model takes WavLM embeddings of speech as the input. The embeddings are processed by two 2D convolutional layers (Conv2D) with 3×3 kernels and rectified linear unit (ReLU) activation functions, followed by two gated recurrent unit (GRU) layers with hidden sizes of 256 and 128, respectively. Batch normalization, dropout (rate = 0.3), and up-sampling (factor of 2) are applied after certain layers as indicated. The output dense layers predict 9 articulatory vocal tract and glottal source variables: 6 tract variables, aperiodicity, periodicity, and fundamental frequency (F0).

Neural network training used the Adam optimizer in conjunction with a learning rate scheduler, starting with an initial learning rate of 5e-4 that reduced when performance plateaued. Grid search revealed that the optimal batch size was 8. To mitigate overfitting, training was terminated if validation performance failed to improve within a patience window of 10 epochs. The speech inversion loss function was defined as a weighted sum of Pearson Correlation ($r$) and Root Mean Squared Error (RMSE) using Equation (2). This loss function ensured that both the absolute error minimization and the correlation of observed and predicted *vocal tract variables* were monitored during training. Through experimentation, the optimal value for $\alpha$ was determined to be 0.8.

$$Loss = \alpha * (1 - r) + (1 - \alpha) * RMSE$$

(Equation 2)

*Model performance*

Pearson's *r* served as the metric quantifying the accuracy of the speech inversion system when predicting the Wisconsin X-ray Microbeam dataset ground-truth tract and source variables from the audio. As shown in Table 3, the proposed model achieved an average $r$ of 0.84 ($\sigma_r = .076$) for all six *vocal tract variables*. This performance corresponded to a 3.67% relative improvement for the same set of speakers over the baseline system reported by Attia et al. (2024). This improvement highlights the efficacy of our approach in speaker-independent articulatory estimation. Key differences in the proposed speech inversion system versus the baseline model are the use of WavLM-Large over HuBERT-Large embeddings and the addition of glottal source variables to the task. Improving performance with the inclusion of source variables is consistent with the work of Siriwardena and Espy-Wilson (2023).

Table 3: *Pearson's* r *Correlation of Ground Truth with Inferred Vocal Tract and Source Variables*

| Model | Embeddings | LA | LP | TTCL | TTCD | TBCL | TBCD | PER | APER | F0 | Mean (STD) Oral TVs |
|---|---|---|---|---|---|---|---|---|---|---|---|
| Attia et al. (2024) Baseline | HuBERT-Large | .8902 | .7142 | .8032 | .9229 | .7361 | .8180 | N/A | N/A | N/A | .8141 (.082) |
| Proposed Speech Inversion System | WavLM-Large | .9123 | .7485 | .8160 | .9457 | .7846 | .8571 | .9165 | .8487 | .6943 | .8440 (.076) |

*Note*: LA = lip aperture, LP = lip protrusion, TTCL = tongue tip constriction location, TTCD = tongue tip constriction degree, TBCL = tongue body constriction location, TBCD = tongue body constriction degree, PER = periodicity, APER = aperiodicity, F0 = fundamental frequency, TV = Articulatory Phonology vocal tract variable. N/A entries reflect that the baseline model did not estimate source variables.

**Dependent variable for research questions: speech inversion tract variable inference**

Each child SSD audio file in the study was processed by the inference model for the adult-trained speech inversion system. This resulted in a time series output matrix for each input utterance. This matrix contained the inferred values of each of the nine *vocal tract* and *source variables* for the length of the input utterance, sampled at 100 Hz. The phone boundaries placed by forced alignment were used to extract the samples from the speech inversion output matrix that were associated with the phone of interest. For the purposes of this study, each phone interval was represented as the dependent variable in the statistical analysis by 6 *vocal tract variable* values, each the average of an individual *vocal tract variable* time series. The *identities* of the *vocal tract variable* signal channel associated with each *vocal tract variable* value were also included as factors in the statistical analyses, as independent variables.

**Statistical analyses**

The statistical analysis code is included as supplemental material. The hypotheses from our first and second research questions were tested by comparing estimated marginal means from linear mixed-effects models. Two separate linear mixed models (one for /ɹ/ targets, one for /s/ targets) were fit to examine the interaction of *vocal tract variable identity* and *phone* on the outcome, *vocal tract variable value*. A random intercept was included to model speaker specific variation in articulation, with repeated audio files clustered within speakers. Model fitting was performed using restricted maximum likelihood, and degrees of freedom were approximated using Satterthwaite's method. Model fitting and parameter estimation was conducted using the *lme4* package in R. The statistical model repeated for /ɹ/ and /s/ is shown in Equation 3, using lme4 notation rather than statistical notation for readability. In this Equation, *tv = vocal tract variable* and *ID* = participant identity. Recall that *phone* refers to the categorical label indicating correct targets, positive controls, and PERCEPT Rating Scale error subtype, which were summarized in Table 2.

$$\text{tv\_value} \sim \text{tv\_identity} * \text{phone} + (1 \mid \text{ID/Utterance}) \qquad \text{Equation (3)}$$

Both models for /ɹ/ and /s/ converged and revealed significant main effects of *vocal tract variable identity* and *phone*, as well as several significant interactions between *vocal tract variable identity* and *phone*. The estimated marginal means from these models were calculated using the R *emmeans* package and are reported in the results. Reverse pairwise contrasts were applied to compare the *vocal tract variable values* for positive control and errored *phones* against the *vocal tract variable values* for correct phones. To control the false discovery rate given the number of comparisons, *p* value adjustment was completed using the method of Benjamini and Hochberg (1995). This method of correction retains the alpha level for significance at .05.

For our third research question, we employed a third linear mixed model to assess whether an utterance's rater-averaged PERCEPT Rating Scale score predicted the magnitude of articulatory difference in rated phones. Because the perceptual rating of a sound and its articulation might be simultaneously influenced by both v*ocal tract variable constriction location* and *constriction degree*, *mean squared articulatory difference* for a given *articulator* was quantified as the mean square difference of *constriction location* and *constriction degree* for errored sounds to the average of perceptually correct sounds, as in Equation 4. Using this method, articulatory differences for a given *articulator* can be aggregated across *phone* error subtypes, even when the differences would be expected to vary in direction. Articulators were defined as *lips*, *tongue tip*, and *tongue body*.

$$articulatory_{msd} = \left(location_{observed} - \overline{location}_{correct}\right)^2 + \left(degree_{observed} - \overline{degree}_{correct}\right)^2$$

Equation 4

This statistical model for this research question did not include utterances unanimously marked as fully correct on the PERCEPT Rating Scale, to avoid biasing the parameter estimates with regard to correct target sounds that would be expected to have the smallest *mean squared articulatory difference* to the average correct target. Positive controls were also not analyzed for this research question, which would be expected to have the largest mean square difference to the average correct target.

A preliminary linear mixed model was fit, again in *lme4* in R, to first examine the interaction of *articulator* and *PERCEPT Rating Scale score* upon *mean squared articulatory difference*. A random intercept was included to model speaker specific variation in articulation, with repeated audio files clustered within speakers. This model converged, so a second model was attempted to analyze the interactions of *target phoneme*, *articulator,* and *PERCEPT Rating Scale score* upon *mean squared articulatory difference.* The normalization achieved through *mean squared articulatory difference* allowed for /ɹ/ and /s/ to be included in the same model, permitting direct statistical comparisons across target phonemes. This model (Equation 5) converged and is reported in the results. In this Equation, TV = *vocal tract variable*, msd = mean squared articulatory difference, PRS = PERCEPT Rating Scale, and ID = participant identity.

tv_msd ~ articulator * PRS score * target phoneme + (1 | ID/Utterances)    Equation (5)

## Results

### Research Questions 1 and 2: Are adult-trained speech inversion *vocal tract variables* sensitive to phonemic and phonetic articulatory differences in child speech for the positive control phonemes, errored phonemes, and correct /ɹ/ or /s/?

These research questions examined if inferred articulation was sensitive to categorical differences between the phones in this study. There were 33 meaningful comparisons in total for our hypotheses: the comparisons between the positive control phonemes and the correct targets, between the errored phones and correct targets, and between the errored phones and positive control phonemes. These comparisons were summarized in Table 1 and the results are summarized in the tables that follow. Recall that we inverted the signs of tongue tract variables relative to Attia et al. (2024). This was done so the Figures that follow would reflect a typical articulatory space, so more positive *constriction location* values would reflect more anterior articulation and more positive *constriction degree* values would reflect higher (i.e., more constricted) articulations.

*Results for /ɹ/*

We evaluated specific hypotheses regarding differences in tract variable values across perceived error subtypes in /ɹ/ productions, using estimated marginal means contrasts ($\Delta\hat{\mu}$) derived from the interaction terms in the fitted linear mixed model. The marginal means contrasts for /ɹ/ hypotheses are summarized in Table 4.

Table 4: *Estimated marginal means for /ɹ/ errored phones and positive control phonemes, compared to correct /ɹ/.*

| Perceptual category | Tract Variable Hypotheses | Positive Control Phoneme or Phone | $\Delta\hat{\mu}$ | SE | *z* ratio | *p* value (adj) | Evidence for Hypothesis? |
|---|---|---|---|---|---|---|---|
| /ɹ/ → [w] | LP is positive compared to correct /ɹ/ | /w/ | 0.085 | 0.007 | 12.5 | <0.0001 | Yes |
| | | /ɹ/ → [w] | 0.017 | 0.006 | 2.7 | 0.054 | No |
| | TTCD is negative compared to correct /ɹ/ | /w/ | -0.30 | 0.007 | -44.0 | <0.0001 | Yes |
| | | /ɹ/ → [w] | -0.16 | 0.006 | -24.2 | <0.0001 | Yes |
| | TBCL is negative compared to correct /ɹ/ | /w/ | -0.24 | 0.007 | -36.0 | <0.0001 | Yes |
| | | /ɹ/ → [w] | -0.19 | 0.006 | -29.4 | <0.0001 | Yes |
| /ɹ/ → [+vocalic] | LP is negative compared to correct /ɹ/ | /ʌ/ | -0.15 | 0.005 | -28.7 | <0.0001 | Yes |
| | | /ɹ/ → [+vocalic] | -0.13 | 0.005 | -23.6 | <0.0001 | Yes |
| | TTCD is negative compared to correct /ɹ/ | /ʌ/ | -0.15 | 0.005 | -28.9 | <0.0001 | Yes |
| | | /ɹ/ → [+vocalic] | -0.14 | 0.005 | -24.8 | <0.0001 | Yes |
| | TBCD is negative compared to correct /ɹ/ | /ʌ/ | -0.28 | 0.005 | -52.3 | <0.0001 | Yes |
| | | /ɹ/ → [+vocalic] | -0.11 | 0.005 | -20.9 | <0.0001 | Yes |

*Note*: The hypotheses comparing error subtypes and paired articulatory endpoint phonemes are reported parenthetically. LP = lip protrusion, TTCD = tongue tip constriction degree, TBCL = tongue body constriction location, TBCD = tongue body constriction degree. *p* values were adjusted based on using the method of Benjamini and Hochberg (1995). This method of correction retains the alpha level for significance at .05.

**Positive control phoneme /w/ and derhotic /ɹ/ → [w] phones**: The distributions of all lip, tongue tip, and tongue body tract variables for positive control phoneme /w/, /ɹ/ → [w] phones, and correct /ɹ/ are shown in Figure 4.

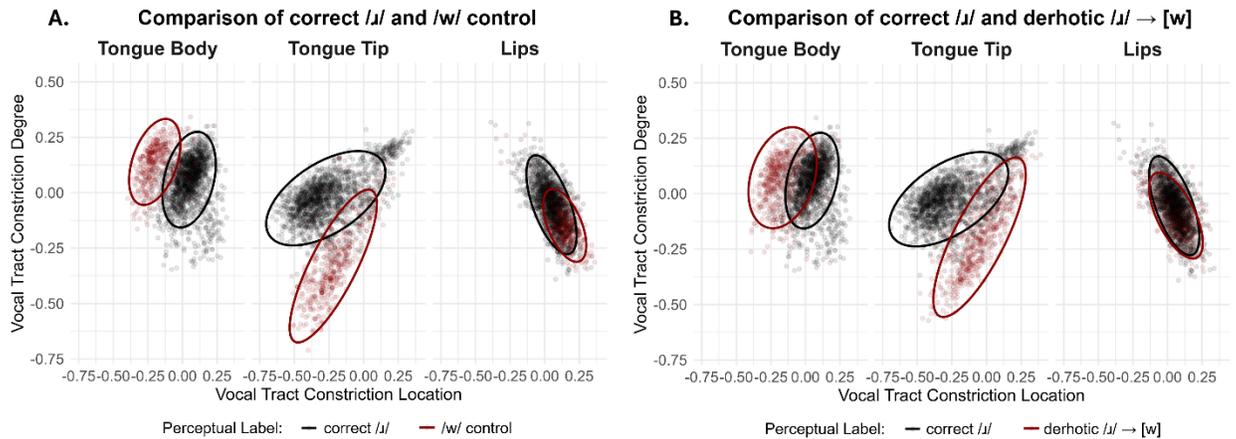

Figure 4: *Data distributions for inferred vocal tract variables for derhotic /ɹ/ → [w] phones and positive control phoneme /w/*. The black distributions correspond to correct /ɹ/ and the red distributions correspond to the positive control phoneme (Panel A) or errored phones (Panel B). In each plot facet, higher values on the x-axis correspond to more anterior locations, and higher values on the y-axis correspond to narrower constrictions. Ellipses correspond to the 95% confidence interval.

We made predictions for how the articulation of the positive control phoneme /w/ would compare to /ɹ/ sounds perceived as correct. There was evidence for all three hypotheses: *lip protrusion* was significantly positive ($\Delta\hat{\mu} = 0.085$, SE = 0.007, $z$ = 12.5, $p_{adj} < 0.0001$), while *tongue tip constriction degree* ($\Delta\hat{\mu}$ = -0.30, SE = 0.007, $z$ = -44.0, $p_{adj} < 0.0001$) and *tongue body constriction location* ($\Delta\hat{\mu}$ = -0.24, SE = 0.007, $z$ = -36.0, $p_{adj} < 0.0001$) were significantly negative, compared to correct /ɹ/.

We made similar predictions for /ɹ/ phones perceived as /ɹ/ → [w] errors. We hypothesized that the lips would be more protruded (i.e., positive *lip protrusion*), that the tongue tip/blade would be lower (i.e., negative *tongue tip constriction degree*), and that the most constricted part of the tongue body would be more posterior (i.e., negative *tongue body constriction location*) compared to /ɹ/ sounds perceived as correct. There was evidence for two of our hypotheses: *tongue tip constriction degree* ($\Delta\hat{\mu}$ = -0.16, SE = 0.006, $z$ = -24.2, $p_{adj}$ < 0.0001) and *tongue body constriction location* ($\Delta\hat{\mu}$ = -0.19, SE = 0.006, $z$ = -29.4, $p_{adj}$ < 0.0001) were significantly negative compared to correct /ɹ/. *Lip protrusion* was positive (i.e., in the expected direction) compared to correct /ɹ/ but the difference did not meet our threshold for significance.

We also hypothesized that productions of the positive control phoneme /w/ would show the same direction of articulatory difference from correct /ɹ/ as derhotic /ɹ/ → [w] phones for each hypothesized contrast, but with /w/ showing a greater magnitude of difference from correct /ɹ/ that /ɹ/ → [w] errors. This was true for each hypothesis: *lip protrusion* ($\Delta\hat{\mu}$ = 0.067, SE = 0.008, $z$ = 8.2, $p_{adj}$ < 0.0001), *tongue tip constriction degree* ($\Delta\hat{\mu}$ = -0.14, SE = 0.008, $z$ = -17.3, $p_{adj}$ < 0.0001), and *tongue body constriction location* ($\Delta\hat{\mu}$ = -.054, SE = 0.008, $z$ = -6.6, $p_{adj}$ < 0.0001) were each significantly greater in the expected direction for /w/ than derhotic /ɹ/ → [w] phones.

**Positive control phoneme /ʌ/ and derhotic /ɹ/ → [+vocalic] phones:** The distributions of all lip, tongue tip, and tongue body tract variables for positive control /ʌ/ phonemes, /ɹ/ → [+vocalic] errors, and correct /ɹ/ are shown in Figure 5.

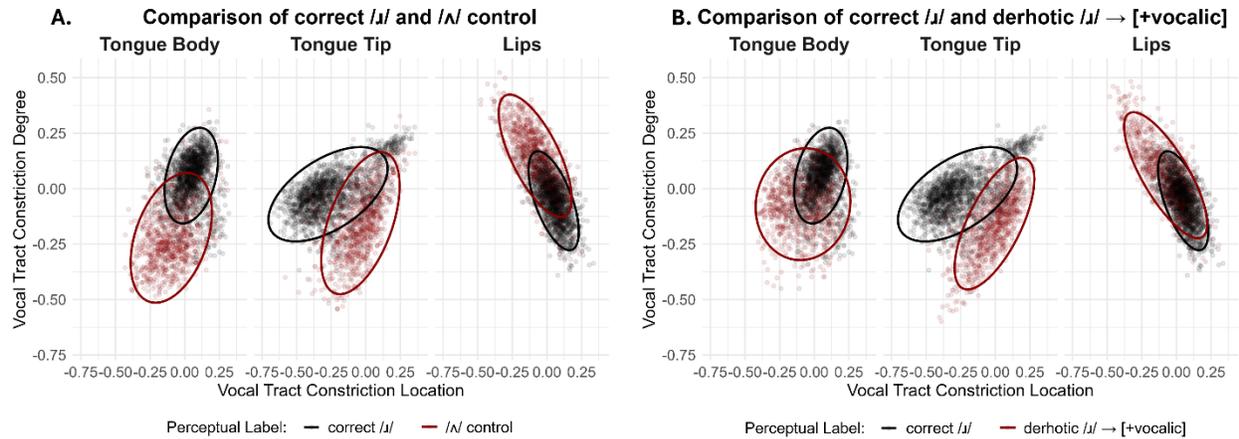

Figure 5: *Data distributions for inferred vocal tract variables for derhotic /ɹ/ → [+vocalic] phones and positive control phoneme /ʌ/.* The black distributions correspond to correct /ɹ/ and the red distributions correspond to the positive control phoneme (Panel A) or errored phones (Panel B). In each plot facet, higher values on the x-axis correspond to more anterior locations, and higher values on the y-axis correspond to narrower constrictions. Ellipses correspond to the 95% confidence interval.

We made predictions for how the articulation of the positive control phoneme /ʌ/ would compare to /ɹ/ sounds perceived as correct. There was evidence for all three hypotheses: LP ($\Delta\hat{\mu}$ = -0.15, SE = 0.005, $z$ = -28.7, $p_{adj} < 0.0001$), *tongue tip constriction degree* ($\Delta\hat{\mu}$ = -0.15, SE = 0.005, $z$ = -28.9, $p_{adj} < 0.0001$), and *tongue body constriction degree* ($\Delta\hat{\mu}$ = -0.28, SE = 0.005, $z$ = -52.3, $p_{adj} < 0.0001$) were each significantly negative compared to correct /ɹ/.

We made similar predictions for /ɹ/ sounds perceived as derhotic /ɹ/ → [+vocalic] phones. We hypothesized that the lips would be less protruded (i.e., negative *lip protrusion*), that the tongue tip/blade would be lower (i.e., negative *tongue tip constriction degree*), and that the tongue body constriction would be looser (i.e., negative *tongue body constriction degree*) compared to /ɹ/ sounds perceived as correct. There was evidence for all three hypotheses: *lip* *protrusion* ($\Delta\hat{\mu}$ = -0.13, SE = 0.005, $z$ = -23.6, $p_{adj}$ < 0.0001), *tongue tip constriction degree* ($\Delta\hat{\mu}$ = -0.14, SE = 0.005, $z$ = -24.8, $p_{adj}$ < 0.0001), and *tongue body constriction degree* ($\Delta\hat{\mu}$ = -0.11, SE = 0.005, $z$ = -20.9, $p_{adj}$ < 0.0001) were each significantly negative compared to correct /ɹ/.

We also hypothesized that productions of the positive control phoneme /ʌ/ would show the same direction of articulatory difference from correct /ɹ/ as derhotic /ɹ/ → [+vocalic] phones for each hypothesized contrast, but with /ʌ/ showing with a greater magnitude of difference from correct /ɹ/ than /ɹ/ → [+vocalic] errors. This was true for each hypothesis: *lip protrusion* ($\Delta\hat{\mu}$ = -0.023, SE = 0.006, $z$ = -3.87, $p_{adj}$ = 0.0011), *tongue tip constriction degree* ($\Delta\hat{\mu}$ = -0.017, SE = 0.006, $z$ = -2.94, $p_{adj}$ = 0.028), and *tongue body constriction degree* ($\Delta\hat{\mu}$ = -0.16, SE = 0.006, $z$ = -26.8, $p_{adj}$ < 0.0001) were each significantly greater in the expected direction for /ʌ/ than /ɹ/ → [+vocalic] errors.

*Results for /s/*

As we did for /ɹ/, we also evaluated specific hypotheses regarding differences in tract variable values across perceived error subtypes in /s/ productions, using estimated marginal means contrasts ($\Delta\hat{\mu}$) derived from the interaction terms in the fitted linear mixed model. The contrasts between these marginal means for /s/ are summarized in Table 5.

Table 5: *Estimated marginal means for /s/ errored phones and positive control phonemes, compared to correct* t /s/.

| Perceptual category | Tract Variable Hypotheses | Positive Control Phoneme or Phone | $\Delta\hat{\mu}$ | SE | $z$ ratio | $p$ value (adj) | Evidence for Hypothesis? |
|---|---|---|---|---|---|---|---|
| /s/ → [s̪] | TTCL is positive compared to correct /s/ | /θ/ | 0.073 | 0.005 | 13.7 | <0.0001 | Yes |
| | | /s/ → [s̪] | 0.051 | 0.006 | 9.1 | <0.0001 | Yes |
| | TBCD is negative compared to correct /s/ | /θ/ | 0.029 | 0.005 | 5.5 | <0.0001 | No |
| | | /s/ → [s̪] | 0.031 | 0.006 | 5.6 | <0.0001 | No |
| /s/ → [sʲ] | TTCL is negative compared to correct /s/ | /ʃ/ | -0.16 | 0.005 | -29.0 | <0.0001 | Yes |
| | | /s/ → [sʲ] | -0.11 | 0.009 | -12.3 | <0.0001 | Yes |
| /s/ → [sˡ] | TTCD is positive compared to correct /s/ | /l/ | -0.16 | 0.009 | -19.1 | <0.0001 | No |
| | | /s/ → [sˡ] | -0.021 | 0.007 | -2.92 | 0.067 | No |
| | TBCD is negative compared to correct /s/ | /l/ | -0.059 | 0.009 | -6.9 | <0.0001 | Yes |
| | | /s/ → [sˡ] | 0.11 | 0.007 | 14.7 | <0.0001 | No |

*Note*: The hypotheses comparing error subtypes and paired articulatory endpoint phonemes are reported parenthetically. TTCL = tongue tip constriction location, TTCD = tongue tip constriction degree, TBCD = tongue body constriction degree, [s̪] = dentalized /s/, [sʲ] = palatalized /s/, and [sˡ] = lateralized /s/. $p$ values were adjusted based on using the method of Benjamini and Hochberg (1995). This method of correction retains the alpha level for significance at .05.

**Positive control phoneme /θ/ and dentalized /s/ → [s̪] phones:** The distributions of all lip, tongue tip, and tongue body tract variables for positive control phoneme /θ/, /s/ → [s̪] errors, and correct /s/ are shown in Figure 6.

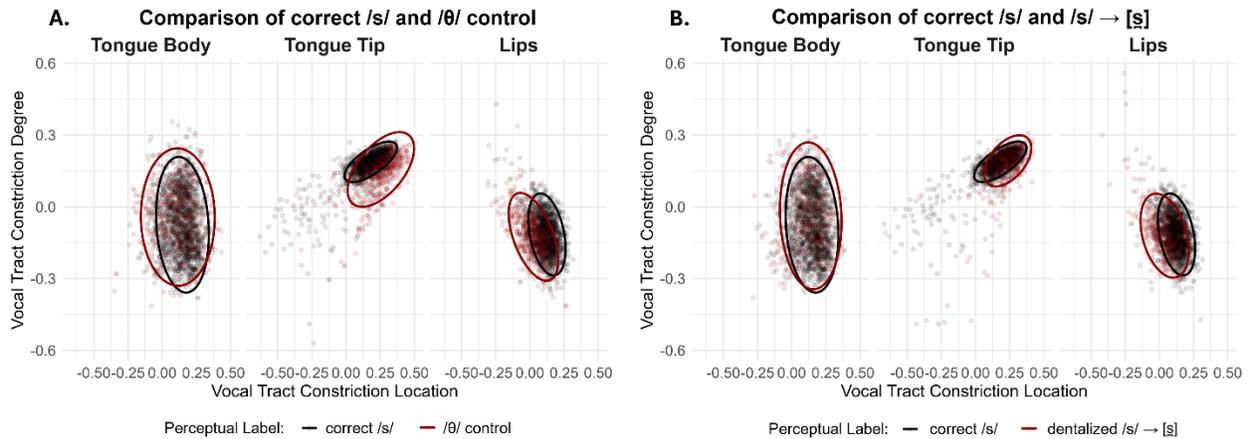

Figure 6: *Data distributions for inferred vocal tract variables for dentalized /s/ → [s̪] phones and positive control phoneme /θ/.* The black distributions correspond to correct /s/ and the red distributions correspond to the positive control phoneme (Panel A) or errored phones (Panel B). In each plot facet, higher values on the x-axis correspond to more anterior locations, and higher values on the y-axis correspond to narrower constrictions. Ellipses correspond to the 95% confidence interval.

We made predictions for how the articulation of the positive control phoneme /θ/ would compare to /s/ sounds perceived as correct. There was evidence for our *tongue tip constriction location* hypothesis ($\Delta\hat{\mu} = 0.073$, SE = 0.005, $z = 13.7$, $p_{adj} < 0.0001$), but not our *tongue body constriction degree* hypothesis, which was in the unexpected direction along with /s/ → [s̪] error.

We made similar predictions for /s/ sounds perceived as dentalized /s/ → [s̪] errors. We hypothesized that the tongue tip/blade would be more anterior (i.e., positive *tongue tip constriction location*) and that the tongue body constriction would be looser (i.e., negative *tongue body constriction degree*) compared to /s/ sounds perceived as correct. There was evidence that *tongue tip constriction location* was more positive than correct /s/ ($\Delta\hat{\mu} = 0.051$, SE $= 0.006$, $z = 9.1$, $p_{adj} < 0.0001$), but we did not find evidence that *tongue body constriction degree* was negative compared to correct /s/. In fact, there was evidence that *tongue body constriction degree* was significantly more positive.

We also hypothesized that productions of the positive control phoneme /θ/ would show the same direction of articulatory difference from correct /s/ as dentalized /s/ → [s̪] phones for each hypothesized contrast, but with /θ/ showing a greater magnitude of difference from correct /s/ than [s̪] errors. We found evidence of that *tongue tip constriction location* ($\Delta\hat{\mu} = 0.022$, SE $= 0.006$, $z = 3.4$, $p_{adj} = 0.016$) was significantly greater in the expected anterior direction for /θ/ than /s/ → [s̪] errors, but this pattern was not seen for *tongue body constriction degree*.

**Positive control phoneme /ʃ/ and palatalized /s/ → [sʲ] phones:** The distributions of all lip, tongue tip, and tongue body tract variables for positive control phoneme /ʃ/, /s/ → [sʲ] errors, and correct /s/ are shown in Figure 7.

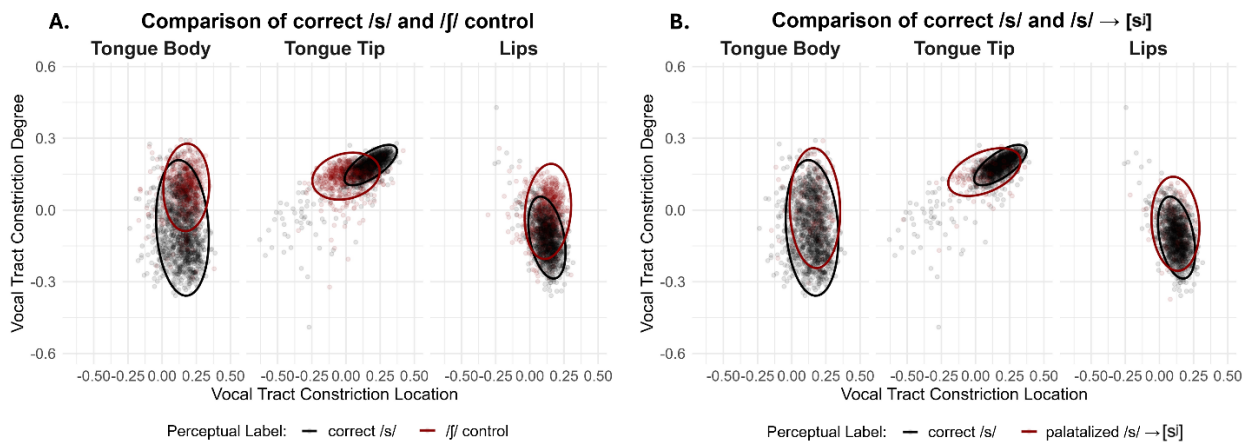

Figure 7: *Data distributions for inferred vocal tract variables for palatalized /s/ → [sʲ] phones and positive control phoneme /ʃ/.* The black distributions correspond to correct /s/ and the red distributions

correspond to the positive control phoneme (Panel A) or errored phones (Panel B). In each plot facet, higher values on the x-axis correspond to more anterior locations, and higher values on the y-axis correspond to narrower constrictions. Ellipses correspond to the 95% confidence interval.

We made predictions for how the articulation of the positive control phoneme /ʃ/ would compare to /s/ sounds perceived as correct. There was also evidence that *tongue tip constriction location* for /ʃ/ was significantly more negative, compared to correct /s/ ($\Delta\hat{\mu}$ = -0.16, SE = 0.005, $z$ = -29.0, $p_{adj}$ < 0.0001).

We made similar predictions for /s/ sounds perceived as backed /s/ → [sʲ] errors. We hypothesized that the tongue tip/blade would be more posterior (i.e., negative *tongue tip constriction location*) compared to /s/ sounds perceived as correct. The analysis supported this ($\Delta\hat{\mu}$ = -0.11, SE = 0.009, $z$ = -12.3, $p_{adj}$ < 0.0001).

We also hypothesized that productions of the positive control phoneme /ʃ/ would show the same direction of articulatory difference from correct /s/ as errored /s/ → [sʲ] phones for each hypothesized contrast, but with /ʃ/ showing a greater magnitude of difference from correct /s/ than [sʲ] errors. The analysis supported that *tongue tip constriction location* was significantly greater in the expected direction for /ʃ/ than errored /s/ → [sʲ] phones ($\Delta\hat{\mu}$ = -0.046, SE = 0.010, $z$ = -4.8, $p_{adj}$ < 0.0001).

**Positive control phoneme /l/ and lateralized /sˡ/ phones:** The distributions of all lip, tongue tip, and tongue body tract variables for positive control phoneme /l/, /s/ → [sˡ] errors, and correct /s/ are shown in Figure 8.

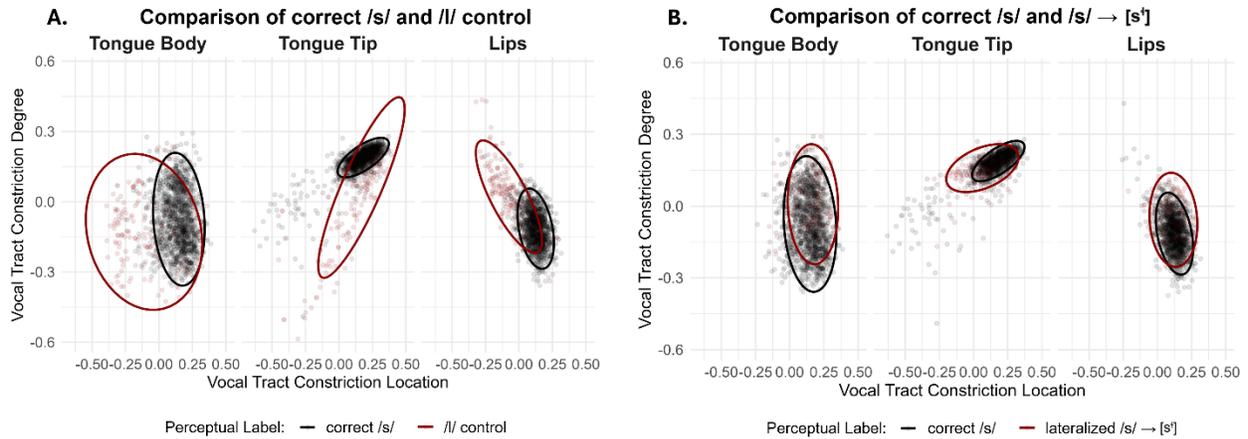

Figure 8: *Data distributions for inferred vocal tract variables for lateralized /s/ → [sˡ] phones and positive control phoneme /l/*. The black distributions correspond to correct /s/ and the red distributions correspond to the positive control phoneme (Panel A) or errored phones (Panel B). In each plot facet, higher values on the x-axis correspond to more anterior locations, and higher values on the y-axis correspond to narrower constrictions. Ellipses correspond to the 95% confidence interval.

We made predictions for how the articulation of the positive control phoneme /l/ would compare to /s/ sounds perceived as correct. There was evidence that *tongue body constriction degree* was significantly more negative in /l/ compared to correct /s/ ($\Delta\hat{\mu}$ = -0.059, SE = 0.009, $z$ = -6.9, $p_{adj}$ < 0.0001), but there was no evidence that *tongue tip constriction degree* was positive in /l/ compared to correct /s/.

We made similar predictions for /s/ sounds perceived as /s/ → [sˡ] errors. We hypothesized that the tongue tip/blade would be higher (i.e., positive *tongue tip constriction degree*) and that the tongue body constriction would be looser (i.e., negative *tongue body* *constriction degree*) compared to /s/ sounds perceived as correct. The analysis did not support either the *tongue tip constriction degree* or *tongue body constriction degree* hypothesis.

We also hypothesized that productions of the positive control phoneme /l/ would show the same direction of articulatory difference from correct /s/ as errored /s/ → [s$^t$] phones for each hypothesized contrast, but with /l/ showing a greater magnitude of difference from correct /s/ than /s$^t$/ errors. The analysis did support this for *tongue tip constriction degree* ($\Delta\hat{\mu}$ = -0.14, SE = 0.010, $z$ = -13.7, $p_{adj}$ < 0.0001), but not *tongue body constriction degree*. In the case of *tongue body constriction degree*, /l/ was indeed more negative (Table 5) but was in a different direction than *tongue body constriction degree* in /s/ → [s$^t$] errors.

## Research Question 3: Does the PERCEPT Rating Scale score correspond to the amount of inferred articulatory difference between errored phones and perceptually correct sounds?

While the analyses for our first and second research questions examined errors categorically, the analysis for our third research question examined whether PERCEPT Rating Scales scores predicted continuous differences in articulation. We evaluated the hypothesis that as PERCEPT Rating Scale scores decreased (i.e., were rated as more incorrect), *mean squared articulatory distance* from the average correct articulation would increase. As shown in supplemental Table S1, linear mixed modeling revealed significant main effects for *articulator*, *PERCEPT Rating Scale score*, and *target phoneme* on *mean squared articulatory difference*. As expected, perceptually incorrect productions rated as closer to fully correct (i.e., higher *PERCEPT Rating Scale scores*) showed smaller overall *mean squared articulatory difference* from the average correct target ($\beta$ = –0.019, SE = 0.004, $t$ = -5.47, $p$ < 0.001). Compared to *lip articulations*, *tongue tip* ($\beta$ = 0.067, SE = 0.007, $t$ = 9.66, $p$ < 0.001) and t*ongue body articulations* ($\beta$ = 0.020, SE = 0.007, $t$ = 2.91, $p$ = 0.004) were associated with significantly higher *mean squared articulatory difference*. *Mean squared articulatory difference* was significantly lower, overall, for errored */s/ phones* than for derhotic /ɹ/ *phones* ($\beta$ = -0.027, SE = 0.01, $t$ = -2.61, $p$ = 0.01).

Two-way interactions between *articulator* and *PERCEPT Rating Scale scores* were not statistically significant, suggesting that the relationship between perceptual accuracy and articulatory difference did not reliably differ across *lip*, *tongue tip*, and *tongue body articulations* in the reference /ɹ/ context. *Tongue tip articulations* showed a significant interaction with *target phoneme* ($\beta$ = –0.080, SE = 0.01, $t$ = -7.14, $p$ < .001), indicating that *mean squared articulatory difference* at the tongue tip was smaller in errored /s/ phones than in derhotic /ɹ/ phones. Similarly, a significant interaction between *PERCEPT Rating Scale* score and *target phoneme* ($\beta$ = 0.010, SE = 0.005, $t$ = 2.33, $p$ = .02) indicated that the expected negative association between *PERCEPT Rating Scale scores* and *mean squared articulatory difference* was weaker for errored /s/ phones than errored /ɹ/ phones.

Significant three-way interactions between *tongue tip articulation*, *PERCEPT Rating Scale score*, and *target phoneme* indicated that the relationship between *PERCEPT Rating Scale Score* and *mean squared articulatory difference* differed across errored /ɹ/ and /s/ phones for *tongue tip articulations*. For *tongue tip articulations* of errored /s/ phones, the significant three-way interaction ($\beta$ = 0.013, SE = 0.006, $t$ = 2.18, $p$ = .029) modifies the expected negative two-way slope between *PERCEPT Rating Scale score* and *mean squared articulatory difference*, resulting in a less negative association for tongue tip in errored /s/ phones than in /ɹ/ phones. No other three-way interactions were significant. These results provide support and contextualization for our hypothesis, indicating that lower scores on the PERCEPT Rating Scale corresponded to higher mean squared differences in *vocal tract constriction location* and *degree*.

This provides evidence that gradient differences in articulation are related to gradient perceptual ratings measured by raters trained on the PERCEPT Rating Scale.

## Discussion

This speech science investigation examined articulatory patterns among speech sound productions from children with SSDs, to demonstrate the association between articulatory kinematic data inferred through speech inversion and expert clinician perceptual ratings facilitated by the PERCEPT Rating Scale. The long-term goal of this research is to motivate the clinical utility of speech inversion inferences and PERCEPT Rating Scale labels in building clinically useful tools.

The first and second research questions examined if speech inversion *vocal tract variable* inferences were sensitive to the category of articulatory differences between perceptually correct speech sounds, subtypes of errored phones, and positive control phonemes. We examined 33 *a priori* hypotheses for /ɹ/ and /s/, the most impacted sounds in chronic SSD. These hypotheses were grounded in articulatory phonetics, imaging of the tongue and vocal tract, vocal tract modeling, and Articulatory Phonology *vocal tract variables*.

For /ɹ/, 17 of 18 hypotheses were supported statistically. Notably, *tongue tip constriction degree, tongue body constriction location,* and *tongue body constriction degree* differentiated derhotic /ɹ/ → [w] and /ɹ/ → [+vocalic] errors from correct /ɹ/. The only unsupported hypothesis for /ɹ/ regarded *lip protrusion* in derhotic /ɹ/ → [w] errors. While moderate lip rounding is considered a salient part of the complex vocal tract configuration for /ɹ/ (Preston, Benway, et al., 2020), these findings suggest that lip configuration has less of an articulatory impact on the perception of derhotic /ɹ/ → [w] than the tongue tip and tongue body. Figure 4 illustrates that, in a clinical setting, there is likely more room for improvement if clinicians perceiving /ɹ/ → [w] errors cued the tongue body rather than the lips. Overall, the findings here suggest that speech inversion inferences can differentiate the tongue dynamics for /ɹ/ → [w] and /ɹ/ → [+vocalic] errors from correct /ɹ/ in articulatory and clinically interpretable ways. Interestingly for /ɹ/, Figures 4 and 5 visually reveal two discernable sub-clusters of tongue tip articulations in correct /ɹ/, and future research can investigate if these sub-clusters reflect "bunched" and "retroflexed" configurations.

For /s/, 7 of 15 hypotheses were supported by the data. In particular, hypotheses related to *tongue tip constriction location* were verified by the data. This may be because /s/ is a coronal phone and the tongue tip is the primary active articulator for coronal sounds. Our hypotheses related to the tongue body, however, were not supported. It may be that *tongue body constriction* variables are less interpretable in cases such as /s/, when the tongue body is not a primary articulator. The statistical models for our third research question also suggest that /s/ uses a smaller working space in the vocal tract than /ɹ/, aligning with the articulatory overlap seen visually between the distributions in Figures 4-8. Also, many of the tongue body constriction degree hypotheses were based on the thought that lateral bracing in correct /s/ would also elevate the midline of the tongue body relative to /s/ errors, but this was not supported by the data and recent accounts do indeed provide evidence that different muscles and their neuromuscular compartments work to provide independent control of different sectors of the midsagittal and lateral tongue body (Wrench, 2024). In all, findings from our first and second research questions may indicate that, in its current form, midsagittally-derived speech inversion provides more interpretable information about /ɹ/, and about the tongue tip in the case of /s/.

The third research question examined if PERCEPT Rating Scale scores corresponded to the amount of articulatory difference between errored sounds and perceptually correct sounds, as measured by speech inversion. Our primary hypothesis was supported: phones with lower PERCEPT Rating Scale scores (i.e., more perceptually incorrect) showed larger mean squared articulatory differences from the average correct target. To our knowledge, this is some of the first evidence that perceptual ratings can quantify not only intermediate acoustic properties but also articulatory proximity to the target sound. This finding extends prior work illustrating how intermediate perceptual ratings in children with /ɹ/-based SSD correspond to acoustically intermediate productions (e.g., Benway et al., 2021), by showing that this relationship also holds for articulatory configurations. This research question also provides evidence for the validity of the PERCEPT Rating Scale and training pipeline to capture articulatory proximity to the target phoneme through perceptual means.

Specifically, our experiments showed that tongue tip and tongue body articulations were associated with significantly higher *mean square articulatory differences* than lip articulations, which is intuitive from an articulatory and clinical perspective. This provides additional evidence that clinical cues for /ɹ/ should not overemphasize the lips. Overall articulatory differences were significantly greater for /ɹ/ targets than for /s/ targets, again suggesting that /ɹ/ involves a greater vocal tract working space than /s/, and that a perceptually correct /s/ might involve a spectral shape achievable through multiple articulatory configurations. This also aligns with prior research indicating that subtle variability in children's place of articulation for /s/ impacts the /s/ spectrum (Munson, 2004). It may be that the smaller vocal tract working space and spectral shape of /s/ interact with speaker specific factors, such as how a perceptually correct /s/ can be produced using either a "tongue tip up" (i.e., apico-alveolar /s/) or "tongue tip down" (i.e., lamino-alveolar /s/) configuration, both of which may differ substantially in articulation despite resulting in similar spectral shapes. Future research can investigate this, but overall, the findings from this research question emphasize that the PERCEPT Rating Scale and training pipeline can provide a perceptual measure of articulatory proximity to a correct target, that is stronger in /ɹ/ than in /s/.

Importantly, it is the case that these speech science questions were answered with *vocal tract variables* that were inferred from a speech inversion deep neural network. This speech inversion neural network was trained and validated on adult articulography data, but the results here add to the growing evidence that adult-trained speech inversion can elucidate relevant information about child speech. Child vocal tracts, however, are smaller and differently shaped than adult vocal tracts (Goldstein, 1980), and it may be that speech inversion trained with child articulography data might further improve the sensitivity of speech inversion for articulatory kinematic analyses of childhood SSD, particularly given the smaller working vocal tract space for /s/. Further research can also investigate the role played by glottal *source variables* in the generalization of adult speech inversion to child speech. It is worth mentioning that speech inversion can be used to drive automated systems that could potentially replicate human clinical judgments for SSD in the future. The long-term vision of this work is to develop tools that support clinicians by enhancing objectivity and efficiency, while preserving the vital role of human expertise and judgment. This aligns with broader principles of responsible clinical AI development, which emphasize transparency, human-in-the-loop systems, and maintaining evidence-based, client-centered clinical practices (Benway & Preston, 2025).

**Limitations**

The limitations of this study also inform our future directions, such as validating speech inversion for child speech and speech sound errors on sounds other than /ɹ/ or /s/. Additionally, tongue root dynamics are important for /ɹ/ (Boyce, 2015), but the available articulography-based datasets to train speech inversion do not contain tongue root information. Our team's ongoing work identifies methodologies to train speech inversion systems that model the tongue root. For /s/, tongue body measures showed complex patterns with regards to our first and second research questions, possibly because the tongue body plays a secondary role in /s/ articulation. These patterns may also be because the current geometric transformations for tongue body *tract variables* can refer to a point anywhere between the midventral to dorsal tongue, depending on the point along this arc minimizing constriction. Our ongoing work aims to improve the interpretability of tongue body *tract variable* location and represent phones by more than one averaged point in time, which may have attenuated some of the lip differences between /ɹ/ and the gliding approximant /w/. Also, selecting /l/ as a positive control phoneme for /s/ → [sˡ] may have confounded lateralization with voicing and manner differences. We plan to continue investigating /s/ → [sˡ] articulation by seeking audio corpora that contain phonemic /ɬ/ (e.g., Welsh, Zulu, or Xhosa).

## Conclusion

This study provides empirical support for using inferred articulatory kinematic data to characterize perceptually rated speech sound errors in children with SSD. Articulatory kinematics were inferred using speech inversion neural networks. Findings demonstrate that inferred articulatory kinematic data were sensitive to both the phonetic category and magnitude of articulatory difference label using the PERCEPT Rating Scale, particularly for productions of /ɹ/. These results support the validation of speech inversion and the PERCEPT Rating Scale for speech science and clinical investigations of childhood SSD.

## Acknowledgements

The authors would like to thank José Ortiz and the Language Diversity Lab at the University of Maryland, College Park for helpful comments on a draft of this manuscript. This research was supported by the NIH (T32DC000046-28, N. Benway, Trainee, C. Espy-Wilson, Mentor, and M. Goupell & C. Carr, PIs; R01DC020959, J. Preston, PI) and the NSF (BCS2141413, C. Espy-Wilson, PI). This study expands upon a preliminary report of five hypotheses to appear in the 2025 *Proceedings of Interspeech*.